\documentclass[11pt]{article}
\usepackage{amssymb}
\usepackage{amsmath}
\usepackage{graphics}
\usepackage{epsfig}
\usepackage{a4wide}
\usepackage{cite}
\usepackage{multirow}
\usepackage{color}
\usepackage{mathrsfs}
\usepackage{slashed}
\usepackage{subfigure}

\DeclareGraphicsExtensions{.eps}
\begin{document}

\title{Hunting for possible Higgs-like boson beyond the Standard Model}
\author{
Xing-Dao Guo$^1$, Jian-Wen Zhu$^{2,3}$\footnote{Corresponding author: zjw0019@mail.ustc.edu.cn},  Ren-You Zhang$^{2,3}$, Shu-Min Zhao$^4$, \\ Wen-Gan Ma$^{2,3}$, and Xue-Qian Li$^5$
\\ \\
{\small $^1$ School of Physics and New Energy, XuZhou University of Technology,}  \\
{\small Xuzhou 221111, Jaingsu, People's Republic of China}  \\
{\small $^2$ State Key Laboratory of Particle Detection and Electronics,}  \\
{\small University of Science and Technology of China, Hefei 230026, Anhui, People's Republic of China}  \\
{\small $^3$ Department of Modern Physics, University of Science and Technology of China,}  \\
{\small Hefei 230026, Anhui, People's Republic of China} \\
{\small $^4$ Department of Physics, Hebei University, Baoding 071002, Hebei, People's Republic of China}  \\
{\small $^5$ School of Physics, Nankai University, Tianjin 300071, People's Republic of China}
}

\date{}
\maketitle
\vskip 10mm

\begin{abstract}
A recent preliminary investigation based on Durgut's report at the American Physical Society site shows a structure at $18.4~ {\rm GeV}$ exists in the invariant mass distribution of $\Upsilon l^+l^- ~ (l = e,\, \mu)$ at the LHC center-of-mass energy of $7$ and $8~ {\rm TeV}$. Its appearance attracts attention of theorists and experimentalists of high energy physics, because it might be a Higgs-like boson of $18.4~ {\rm GeV}$ which would serve as a signal of the new physics beyond the Standard Model.  We have carried out computations on the corresponding quantities (production and decay rates) based on quantum field theory and compared the results with experimental data. Our numerical results do not support the assertion that the $18.4~ {\rm GeV}$ peak corresponds to a neutral $0^{++}$ boson which decays into $\Upsilon l^+l^-$. Much further works (both experimental and theoretical) are badly needed.
\end{abstract}

\vfill \eject
\baselineskip=0.32in
%%%%%%%%%%%%%%%%%%%%%%%%%%%%%%%%%%%% Introduction %%%%%%%%%%%%%%%%%%%%%%%%%%%%%%%%%%%%%%%%%
\makeatletter      % '@' is now a normal "letter" for TeX
\@addtoreset{equation}{section}
\makeatother       % '@' is restored as a "non-letter" character for TeX
\vskip 5mm
\renewcommand{\theequation}{\arabic{section}.\arabic{equation}}
\renewcommand{\thesection}{\Roman{section}.}
\newcommand{\nb}{\nonumber}
%%%%%%%%%%%%%%%%%%%%%%%%%%%%%%%%%%%%%%%%%%%%%%%%%%%%%%%%%%%%%%%%%%%%%%%%%%%%%%%%%%%%%%%%%%%
%\vskip 5mm

\section{Introduction}
\par
The $125~ {\rm GeV}$ Higgs boson was found by the ATLAS and CMS collaborations in 2012, which pastes the last brick of the Standard Model (SM). While physicists were celebrating the success of the SM, several problems emerged in front of us. Definitely we all know by bottom of hearts that the SM by no means is the final theory of the Nature but only an effective one at the concerned energy scale. Its loopholes, such as the naturalness of Higgs boson and the vacuum stability, all compose serious challenge to our knowledge. Moreover, some experimental phenomena show slight deviations from the SM predictions which reveal traces of new physics beyond SM (BSM). Indeed, the target of high energy society is searching for BSM and it especially is the task of LHC which historically succeeded in the SM Higgs boson discovery. Among all possible signals of BSM, the most favorable and significant signal of BSM is the existence of a new Higgs-like boson(s) which is predicted by many new models about BSM.

\par
Recently, a preliminary investigation based on the report by S. Durgut, which is available at the American Physical Society (APS) site \cite{Durgut}, shows a structure around $18.4~ {\rm GeV}$ in four-lepton final state by using the four-lepton events collected during the LHC Run I stage. Below, we just refer the report as D-report. After carefully analysis, they conclude that an enhancement exists at around $18.4~ {\rm GeV}$ in the invariant mass distribution of $\Upsilon l^+l^- ~ (l = e,\, \mu)$ \cite{Durgut,Yi:2018fxo,phdpaper}.

\par
The enhancement is conjectured to occur through the process $pp \rightarrow \Upsilon l^+ l^- \rightarrow \mu^+\mu^-l^+l^-$. Namely if $\Upsilon l^+l^-$ comes from a unique enhancement, it would be a neutral boson $\phi$, which is mainly produced via gluon-gluon fusion, and then sequently decays as $\Upsilon\Upsilon^* \rightarrow \mu^+\mu^- l^+l^-$, where $\Upsilon^*$ might be off-mass-shell due to the energy constraint. The mass of the new enhancement, if it indeed exists, is a few hundreds of MeV lower than the total mass of a $\Upsilon$ pair. By analyzing the datasets collected by the CMS detector at a center-of-mass energy of $7~ {\rm TeV}$ and $8~ {\rm TeV}$ with an integrated luminosity of $25.6~ {\rm fb}^{-1}$ \cite{phdpaper}, and taking the kinematic requirements of $p_{T,l} > 2.0~ {\rm GeV}$ and $\left| \eta_{l} \right| < 2.4$ on the final-state leptons \cite{Durgut,phdpaper}, the experimenters observed a peaking structure. By their rigorous analysis, the peak is located at $18.4 \pm 0.1(stat.) \pm 0.2(syst.)~ {\rm GeV}$ for the $\Upsilon \mu^+\mu^-$ and $\Upsilon e^+e^-$ channels combined; event numbers are $44 \pm 13$ and $35 \pm 13$ for the $\Upsilon \mu^+\mu^-$ and $\Upsilon e^+e^-$ channels, respectively, and its significance is $3.6$ standard deviation after taking into account the look-elsewhere-effect \cite{Durgut,phdpaper}. Moreover, the ANDY collaboration also claims that a significant peak at $m = 18.12 \pm 0.15(stat.) \pm 0.6(syst.)~ {\rm GeV}$ in the dijet mass distribution is observed in Cu+Au collision at $\sqrt{s_{NN}} = 200~{\rm GeV}$ at the Relativistic Heavy Ion Collider \cite{Bland:2019aha}. This result is in good agreement with the four-lepton signal observed at the $7~ {\rm TeV}$ and $8~ {\rm TeV}$ LHC by the CMS collaboration \cite{Durgut,phdpaper}.

\par
Because this enhancement is close to the sum of the masses of four bottom quarks, thus some authors consider it to be a $bb\bar{b}\bar{b}$ tetraquark state with a mass in the range of $18.4~ {\rm GeV} < m_{X_{bb\bar{b}\bar{b}}} < 20.3~ {\rm GeV}$ \cite{Berezhnoy:2011xn,Wu:2016vtq,Chen:2016jxd,Bai:2016int,Anwar:2017toa,Richard:2017vry,Wang:2017jtz}. The authors of Ref.\cite{Karliner:2016zzc} show that $\sigma(pp \rightarrow X_{bb\bar{b}\bar{b}}[0^{++}] \rightarrow 4l)\leqslant 4~ {\rm fb}$ at $\sqrt{s} = 13~ {\rm TeV}$ and $\leqslant 2~ {\rm fb}$ at $\sqrt{s} = 8~ {\rm TeV}$. In Ref.\cite{Esposito:2018cwh} the authors emphasize that their work was motivated by the peak at $18.4~ {\rm GeV}$ based on the tetraquark hypothesis, but the numerical result shows that the partial width for $X_{bb\bar{b}\bar{b}} \rightarrow \Upsilon\mu^+\mu^-$ is too small to tolerate the data currently observed at the LHC. In Ref.\cite{Becchi:2020mjz} the authors calculate the decay width for $X_{bb\bar b\bar b} \rightarrow \Upsilon l^+ l^-$ and they prefer $X_{bb\bar{b}\bar{b}}$ to be a $2^{++}$ tetraquark state rather than a $0^{++}$ bound state. Furthermore, the authors of Ref.\cite{Hughes:2017xie} believe that a $bb\bar{b}\bar{b}$ tetraquark should lie above the lowest noninteracting bottomonium-pair threshold.

\par
Because the mass is just a bit below the sum of two $\Upsilon$ bosons, being driven by the expectation of searching for a BSM Higgs-like boson, it is tempted to conjecture the newly observed enhancement to be a $0^{++}$ fundamental boson. In this work, our purpose is to check if the idea could be tolerated by the experimental observation. To serve this goal, we calculate the production rate of $\Upsilon l^+l^-$ at the LHC by assuming the peak observed in experiment to be real, and then estimate the full contribution from the $18.4~ {\rm GeV}$ structureless BSM boson $\phi(18.4)$ as well as the corresponding SM background. In this paper, the $0^{++}$ enhancement $\phi(18.4)$ is assumed as a BSM Higgs-like boson with mass around $18.4~ {\rm GeV}$. It should be noted that if $\phi(18.4)$ were indeed a Higgs-like boson, and its width is large enough,  a threshold effect would induce an asymmetric peak in the invariant mass spectrum of $\Upsilon l^+l^-$ which could be observed in experiments. Thus we estimate the possibility by numerically calculating the production of $\Upsilon l^+l^-$. We eventually find that the production rate induced by the Higgs-like boson $\phi(18.4)$ is too small to be observed in the LHC with presently available experimental condition. It means that if we deliberately postulate a large width for $\phi(18.4)$, the contribution of the supposed BSM model may generate an experimentally observed peak around $18.4~ {\rm GeV}$, however the data says no.

\par
Discovering new physics beyond SM should begin with looking for a new extra Higgs-like boson(s), this strategy is commonly accepted by both experimentalists and theorists in the high energy physics society. So far by now, many BSM models predict various kinds of  new Higgs-like bosons (for example, neutral, charged, $\mathcal{CP}$-odd or $\mathcal{CP}$-even etc., even doubly-charged bosons). Unfortunately, none of them was found in present experiments so far. When looking back, we find that almost all the particles predicted by those BSM models are much heavier than the SM scale, namely it varies from  few hundreds of GeV to few hundreds of TeV. Such BSM particles cannot be produced in present experimental facilities. On other aspect, there does not exist a principle forbidding the existence of lighter BSM particles. For example, the two-Higgs-doublet-model may predict a new Higgs-like boson with a mass of $28~ {\rm GeV}$ \cite{Cici:2019zir}. By the general method adopted for searching TeV-scale particles at LHC, alternatively, we, in this work, explore a new  Higgs-like boson at low energy regions. As a common sense the strategy can be traced back from our experience gained at lepton colliders, such as BES, Belle, etc. For example, in the scattering process $e^+e^- \rightarrow J/\psi \rightarrow \text{{\it final products}}$, the resonance ($J/\psi$) overwhelmingly dominates the portal, while the direct production just provides a continuous background. For the same cause, a direct production of four leptons from the gluon-gluon fusion at the proton-proton collider, i.e., $gg \rightarrow \Upsilon \Upsilon^* \rightarrow \mu^+\mu^- l^+l^-$ \cite{Li:2009ug,Qiao:2009kg} where $\Upsilon^*$ might be off-mass-shell, should just generate a background. If a medium Higgs-like boson $\phi(18.4)$ indeed exists, it induces the portal of $\Upsilon \Upsilon^* \rightarrow \Upsilon l^+l^-$, a peak would appear in the invariant mass spectrum of $\Upsilon l^+l^-$. With this assertion, we numerically calculate the contribution induced by the BSM Higgs-like boson $\phi(18.4)$ to $pp \rightarrow \Upsilon \Upsilon^* \rightarrow \Upsilon l^+l^-$ at the LHC. Comparing our numerical results with those in the D-report \cite{Durgut}, we find that the assumption that the observed peak in the $\Upsilon l^+l^-$ mass spectrum originates from a BSM Higgs-like boson decay should be ruled out.

\par
This work is organized as follows. After this introduction, in section II, we present our analytical calculation for $\Upsilon l^+l^-$ production at the LHC in the framework of a BSM model, in which we assume that the interaction of the BSM Higgs-like boson with SM particles is in analogue to that of the SM Higgs boson. In section III, we numerically evaluate all corresponding quantities and illustrate the invariant mass distribution of the final state. The last section is devoted to our conclusion and a brief discussion.

\section{Analytical calculation for $pp \rightarrow \Upsilon l^+l^-$}
At the LHC, $\Upsilon l^+l^-$ is mainly produced via gluon-gluon fusion \cite{Georgi:1977gs,Anastasiou:2002yz}, i.e.,
\begin{eqnarray}
\sigma[pp \rightarrow \Upsilon l^+l^-] = \int dx_1 dx_2\, f(x_1, \mu_F)\, f(x_2, \mu_F)\, \hat{\sigma}[gg \rightarrow \Upsilon l^+l^-],
\label{eqn1}
\end{eqnarray}
where $f(x, \mu_F)$ is the gluon distribution function in proton, $\mu_F$ is the  factorization scale, while other production channels are neglected.

\par
The contribution of the BSM Higgs-like boson comes from the Breit-Wigner  propagator $\dfrac{1}{p^2-m_{\phi}^2 + i m_{\phi} \Gamma_{\phi}}$, where $p$ is the four-momentum flowing through the intermediate BSM Higgs-like boson. If we do not consider the interference with the SM background and neglect the $t$- and
$u$-channel Feynman diagrams induced by the BSM Higgs-like boson, this contribution will be proportional to the square of the Breit-Wigner propagator $\dfrac{1}{( s - m_{\phi}^2 )^2 + m_{\phi}^2 \Gamma_{\phi}^2}$, where $s = (k_1+k_2)^2$ and $k_i~ (i = 1, 2)$ are the four-momenta of the two initial-state gluons. When $s$ is close to $m_{\phi}^2$, this factor turns into $\dfrac{1}{m_{\phi}^2 \Gamma_{\phi}^2}$ and a resonance would peak up from the
background. However, if $s$ is far away from $m_{\phi}^2$ (below or above), the contribution of $\phi$ would be drowned into the background and no peak can be seen. In our case, $18.4~ {\rm GeV}$ is slightly below the threshold of $2 m_{\Upsilon}$. However, since its position is not too far from the threshold value and it possesses a relatively large width, the resonance effect still can manifest itself in the invariant mass spectrum of $\Upsilon$ pair at the threshold. In one aspect, the mass of $\phi$ cannot be larger than $2 m_{\Upsilon}$, otherwise a peak at the $\Upsilon$ pair invariant mass spectrum would be seen, but no such peak was experimentally observed.

\par
To evaluate the contribution of the supposed BSM Higgs-like boson $\phi$ of $18.4~ {\rm GeV}$ to the $\Upsilon l^+l^-$ production at the LHC, we write up the complete expression where the Breit-Wigner propagator of $\phi$ with a width observed in the concerned experiment would induce the peak in the $\Upsilon l^+l^-$ invariant mass spectrum. Later our numerical results show that one only needs to account the contribution of the resonance above the threshold of $2 m_{\Upsilon}$. Indeed, because $18.4~ {\rm GeV}$ is smaller than 2$m_{\Upsilon}$, $\phi$ cannot be on its mass-shell for two on-shell $\Upsilon$s, while the production rate for $pp \rightarrow \phi \rightarrow \Upsilon\Upsilon^* \rightarrow \Upsilon l^+l^-$ is very tiny and can be neglected. Due to the extremely small decay width of $\Upsilon$ ($\Gamma_{\Upsilon} \sim 50~ {\rm keV}$ and $\Gamma_{\Upsilon} \ll \Gamma_{\phi}$), the parton-level cross section for the production of $\Upsilon l^+l^-$ via gluon-gluon fusion can be written as
\begin{eqnarray}
\hat{\sigma}[gg \rightarrow \Upsilon l^+l^-] \simeq \hat{\sigma}[gg \rightarrow \Upsilon \Upsilon]
\times
2 \, Br(\Upsilon \rightarrow l^+l^-).
\end{eqnarray}
The cross section $\hat{\sigma}[gg \rightarrow \Upsilon  \Upsilon]$ is given by
\begin{eqnarray}
\hat{\sigma}[gg \rightarrow \Upsilon  \Upsilon] = \int d \Omega \, \left| \mathcal{M}_{SM} + \mathcal{M}_{\phi} \right|^2,
\label{sigmahat}
\end{eqnarray}
where $\mathcal{M}_{SM}$ and $\mathcal{M}_{\phi}$ represent the Feynman amplitudes in the SM  and induced by the BSM Higgs-like boson $\phi$, respectively. It is noted that the above formula is a general expression where we do not specially require the intermediate boson $\phi$, if it indeed exists in the nature, to be real or virtual. The production of $\Upsilon$ pair via gluon-gluon fusion at hadron colliders has been much investigated in the framework of the SM \cite{Li:2009ug,Qiao:2009kg}. The $31$ Feynman diagrams for $gg \rightarrow \Upsilon \Upsilon$ in the SM can be created with the help of {\sc FeynArts} \cite{Hahn:2000kx} package. We also calculate this process with the same input parameters as given in Ref.\cite{Li:2009ug} for comparison, and find that our numerical result for the production cross section at the $14~ {\rm TeV}$ LHC is in good agreement with the corresponding one of Ref.\cite{Li:2009ug} within a tolerable calculation error. Then we step on to calculate the quantities concerning the new Higgs-like boson.

\par
\begin{figure}[htbp]
    \centering
        \subfigure[]{
%            \label{fig:subfig:a}
            \includegraphics[scale=0.35]{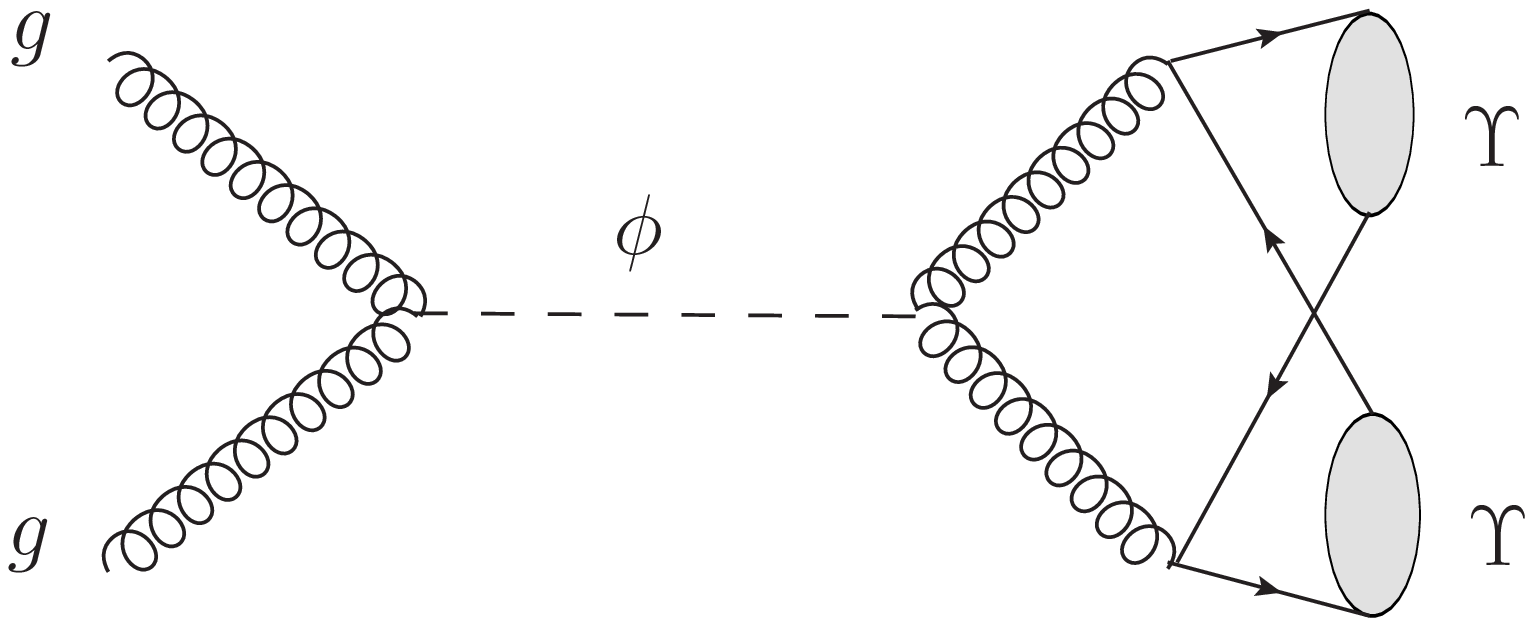}}
%        \hspace{0.08cm}
        \subfigure[]{
%            \label{fig:subfig:b}
            \includegraphics[scale=0.35]{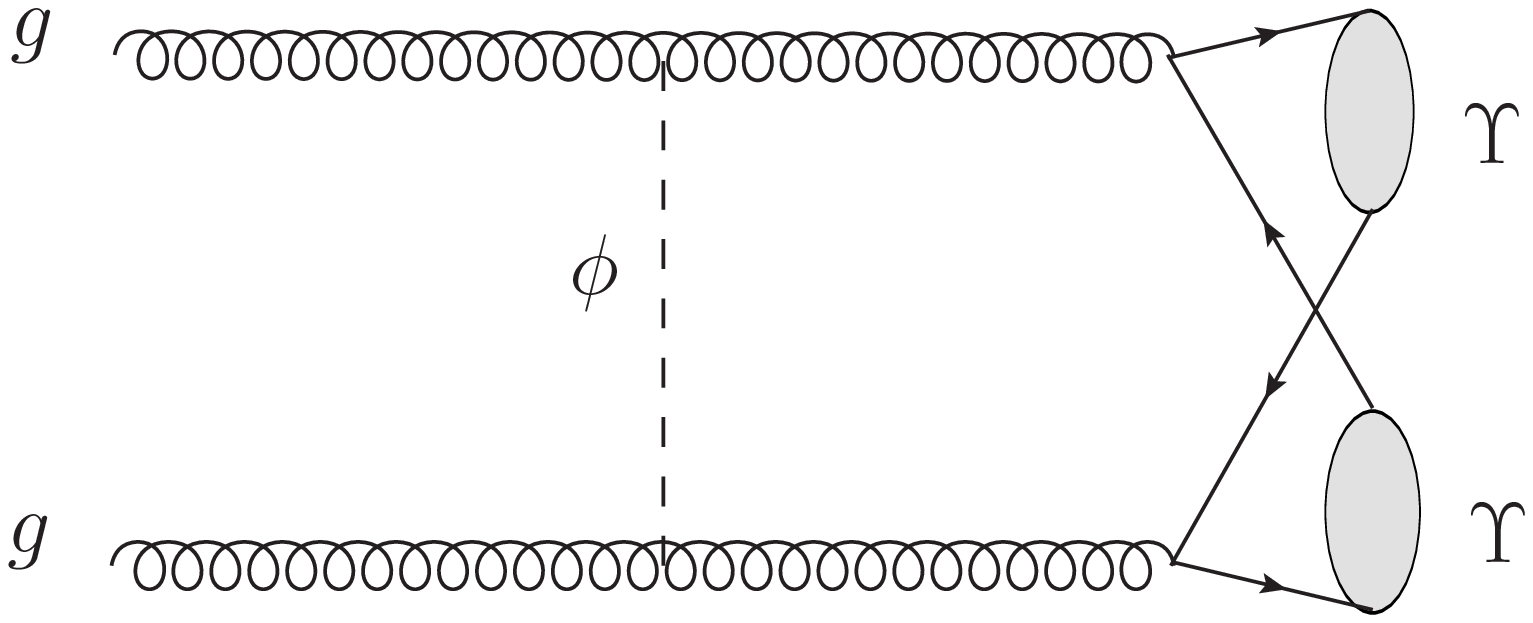}} \\
%        \hspace{0.8cm}
        \subfigure[]{
%            \label{fig:subfig:c}
            \includegraphics[scale=0.35]{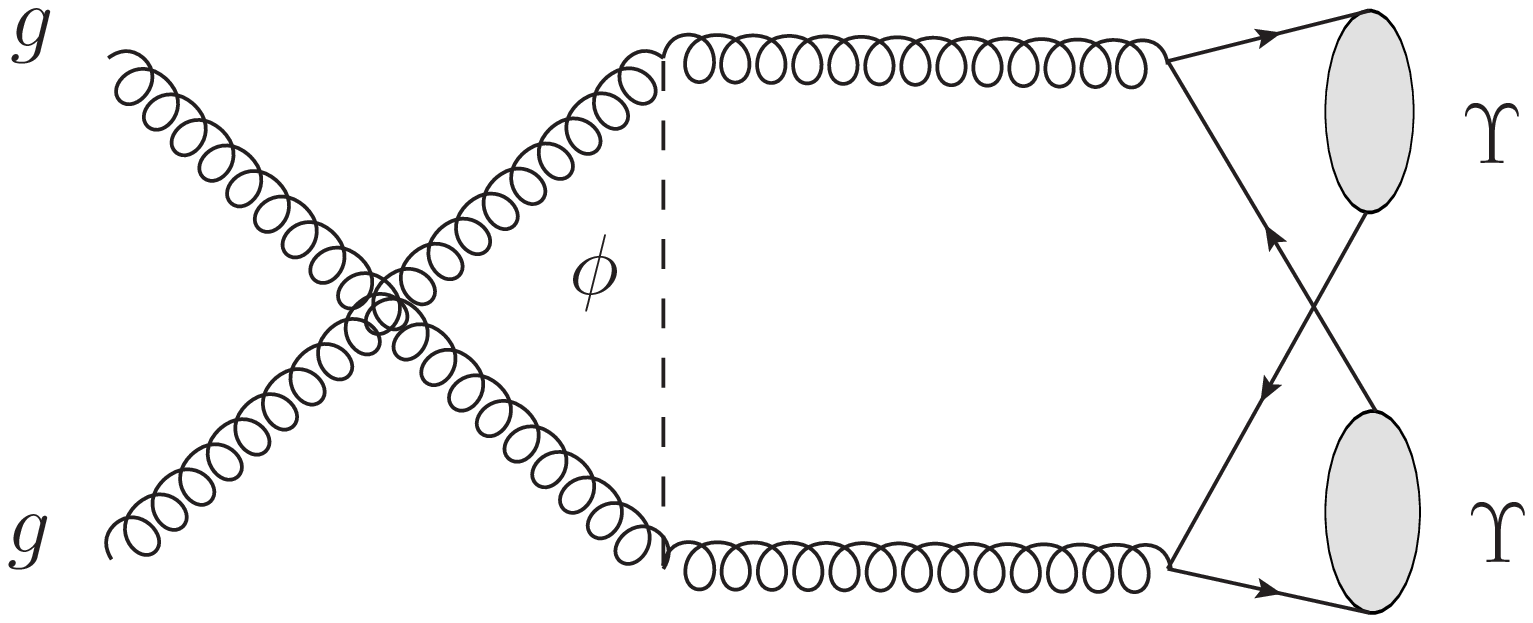}}
\caption{Feynman diagrams for $gg \rightarrow \Upsilon\Upsilon$ induced by the BSM Higgs-like boson $\phi$ via the $gg\phi$ effective coupling.}
\label{fig1}
\end{figure}
\begin{figure}[htbp]
    \centering
        \subfigure[]{
%            \label{fig:subfig:a}
            \includegraphics[scale=0.35]{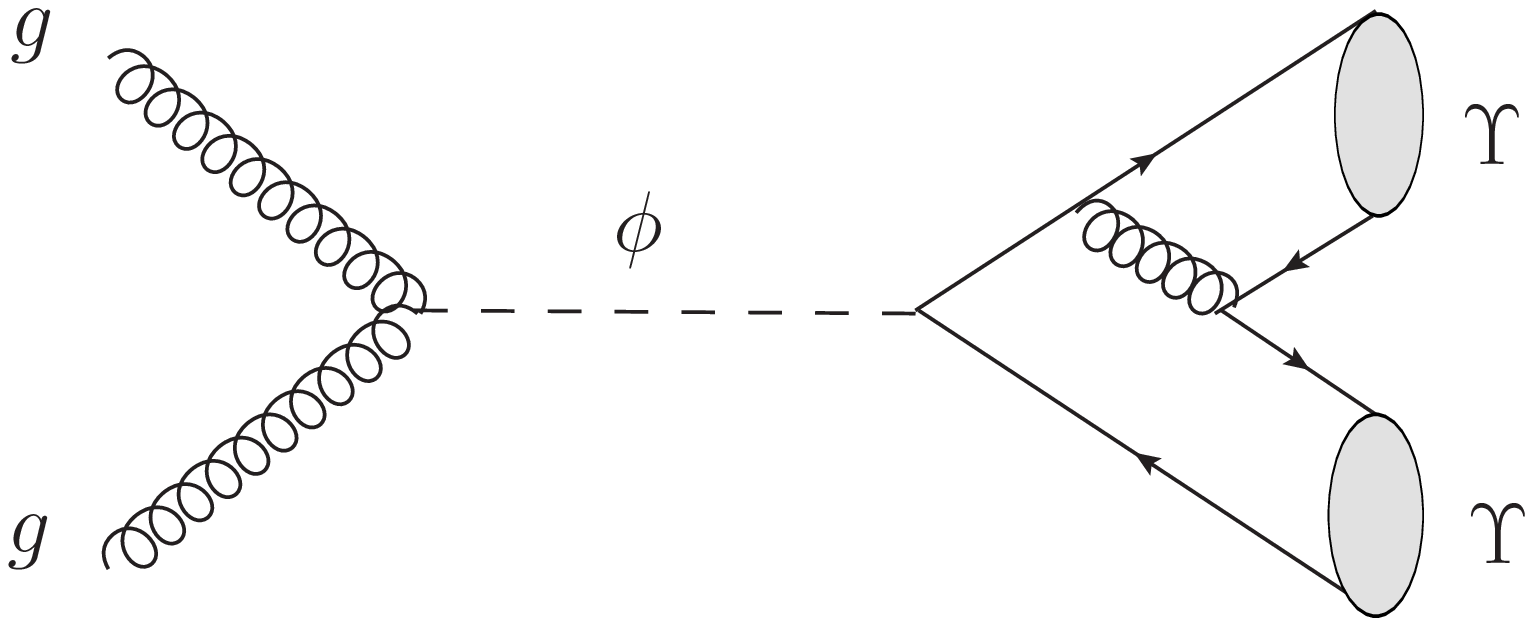}}
        \subfigure[]{
%            \label{fig:subfig:b}
            \includegraphics[scale=0.35]{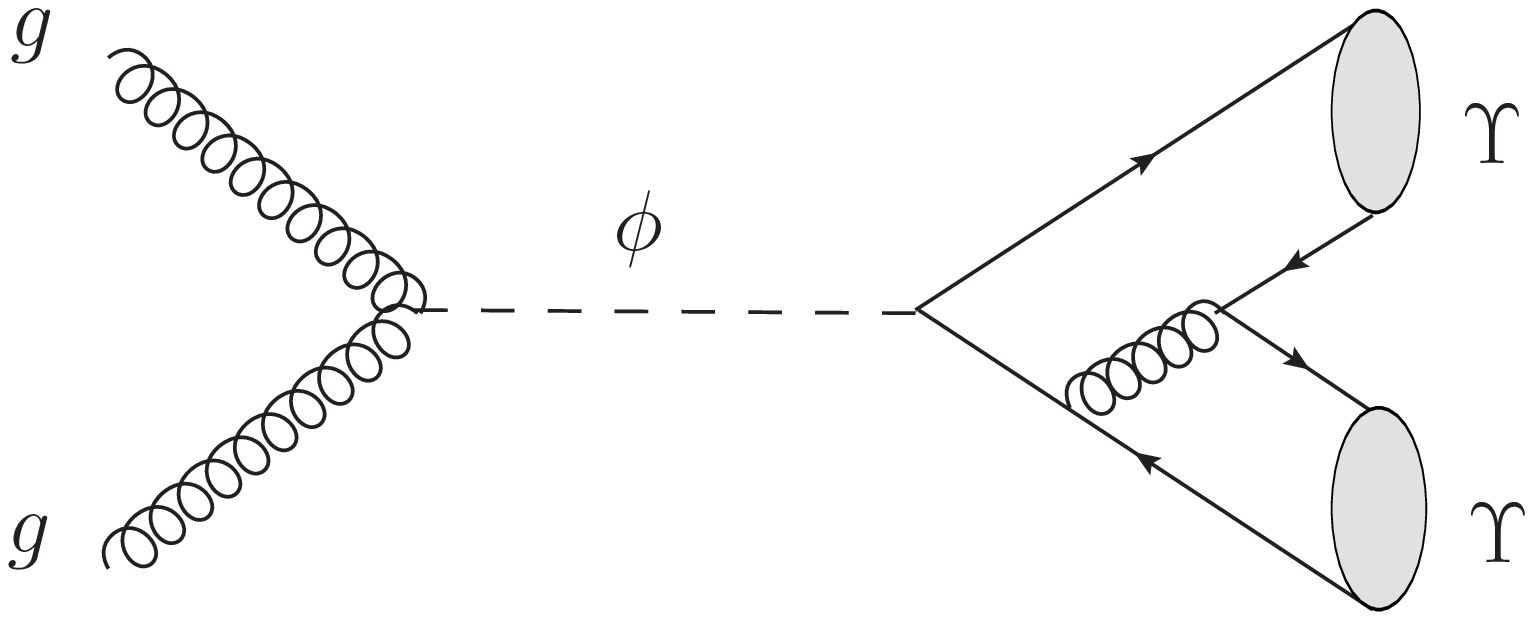}} \\
        \subfigure[]{
%            \label{fig:subfig:c}
            \includegraphics[scale=0.35]{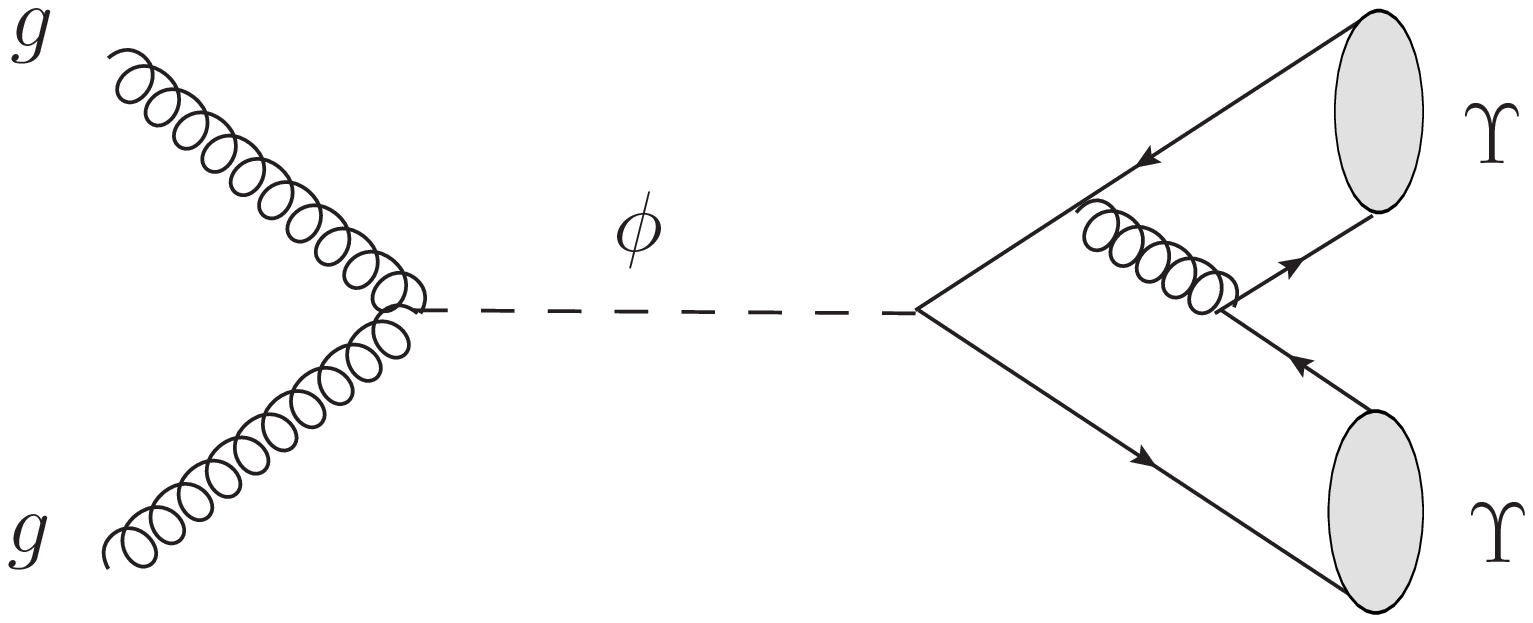}}
        \subfigure[]{
%            \label{fig:subfig:c}
            \includegraphics[scale=0.35]{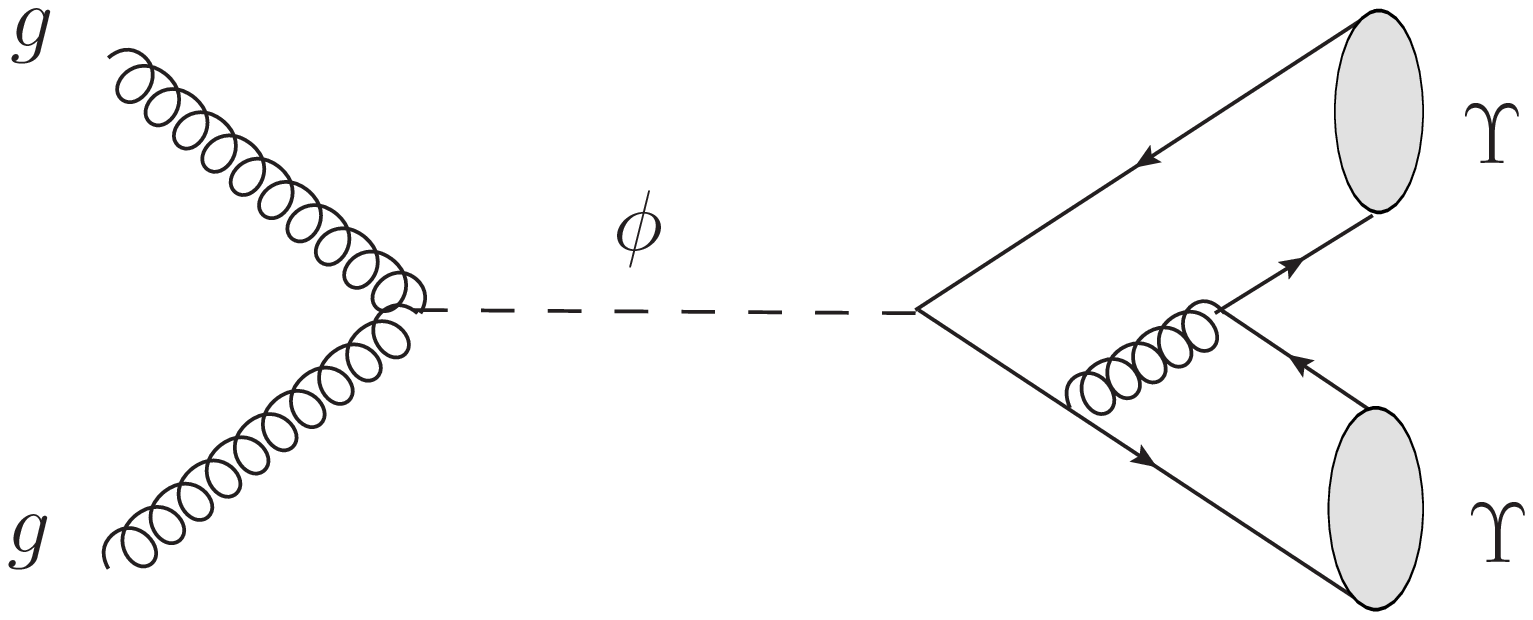}}
\caption{Feynman diagrams for $gg \rightarrow \Upsilon\Upsilon$ induced by the BSM Higgs-like boson $\phi$ via both $gg\phi$ and $b\bar{b}\phi$ effective couplings.}
\label{fig2}
\end{figure}

As for the BSM contribution from the new Higgs-like boson $\phi$, we consider only the $gg\phi$ and $b\bar{b}\phi$ effective couplings for the Higgs-like boson. The Feynman diagrams induced by the BSM Higgs-like boson $\phi$ can be classified into two categories. The diagrams in Fig.\ref{fig1} are independent of the $b\bar{b}\phi$ coupling, while all the diagrams in Fig.\ref{fig2} depend on both $gg\phi$ and $b\bar{b}\phi$ couplings.
%%%%%%%%%%%%%%%%%%%%% Fig. 1 %%%%%%%%%%%%%%%%%%%%%%%%
%%%%%%%%%%%%%%%%%%%%%%%%%%%%%%%%%%%%%%%%%%%%%%%%%%%%%
%%%%%%%%%%%%%%%%%%%%% Fig. 2 %%%%%%%%%%%%%%%%%%%%%%%%
%%%%%%%%%%%%%%%%%%%%%%%%%%%%%%%%%%%%%%%%%%%%%%%%%%%%%

\par
Following Refs.\cite{Anastasiou:2015ema,Spira:2016ztx}, the effective coupling of the BSM Higgs-like boson to two gluons can be written as
\begin{eqnarray}
\label{VggH}
\mathcal{C}_{gg\phi}^{\mu \nu}(k_1,\, k_2)
=
-i\, \dfrac{g_{gg\phi}(\mu_R)}{m_{\phi}} \left[ 4k_1 \cdot k_2 \Big( g^{\mu\nu}-\frac{k_1^\nu k_2^\mu}{k_1\cdot k_2} \Big) \right],
\end{eqnarray}
where $k_1,~ k_2$ and $\mu,~ \nu$ are the four-momenta and Lorentz indices of the two gluons, respectively,  $g_{gg\phi}(\mu_R)$ is a dimensionless effective running coupling constant, and $\mu_R$ is the renormalization scale. It is reasonable to assume that the evolution of the effective coupling constant $g_{gg\phi}$ is the same as that of the QCD $\alpha_s$, i.e., $\dfrac{g_{gg\phi}(\mu_R)}{\alpha_s(\mu_R)}$ is independent of $\mu_R$. Thus, we obtain the quark-level amplitude for the Feynman diagrams in Fig.\ref{fig1} as
\begin{eqnarray}
\widetilde{\mathcal{M}}_{\phi}^{(g)}
=
-i \frac{4 \pi \alpha_s(\mu_R) \lambda_{{\rm color}}}{(p_1 + q_2)^2 (p_2 + q_1)^2}\,
\epsilon_\mu(k_1) \epsilon_\nu(k_2)\,
{\rm Tr} \Big[ v(p_2) \bar{u}(p_1) \gamma_{\alpha} v(q_2) \bar{u}(q_1) \gamma_{\beta} \Big]
\Big(
\mathcal{S} + \dfrac{\mathcal{T} + \mathcal{U}}{8}
\Big),
\end{eqnarray}
where $\lambda_{{\rm color}} \equiv \delta^{ab} {\rm Tr}(T^aT^b) = 4$, and $\mathcal{S}$, $\mathcal{T}$ and $\mathcal{U}$ are given by
\begin{eqnarray}
&&
\mathcal{S} =
\mathcal{C}^{\mu\nu}_{gg\phi}(k_1,\, k_2)\,
\mathcal{C}^{\alpha\beta}_{gg\phi}(p_1+q_2,\, p_2+q_1)
\left/
\Big[ (k_1 + k_2)^2 - m_{\phi}^2 + i m_{\phi} \Gamma_{\phi} \Big]
\right.,
 \\
&&
\mathcal{T} =
\mathcal{C}^{\mu\alpha}_{gg\phi}(k_1,\, p_1+q_2)\,
\mathcal{C}^{\nu\beta}_{gg\phi}(k_2,\, p_2+q_1)
\left/
\Big[ (p_1 + q_2 - k_1)^2 - m_{\phi}^2 + i m_{\phi} \Gamma_{\phi} \Big]
\right.,
 \\
&&
\mathcal{U} =
\mathcal{T}\,\Big|_{k_1 \leftrightarrow k_2,~ \mu \leftrightarrow \nu}\,.
\end{eqnarray}
The effective coupling of the BSM Higgs-like boson to the bottom quarks is parameterized as
\begin{eqnarray}
\label{VbbH}
\mathcal{C}_{b\bar{b}\phi} = -i\, g_{b\bar{b}\phi}(\mu_R).
\end{eqnarray}
We assume that the evolution of the effective coupling constant $g_{b\bar{b}\phi}$ is the same as that of the bottom-quark $\overline{{\rm MS}}$ running mass $\overline{m}_b(\mu_R)$ \cite{Bednyakov:2016onn,Tanabashi:2018oca}, i.e., $\dfrac{g_{b\bar{b}\phi}(\mu_R)}{\overline{m}_b(\mu_R)}$ is independent of $\mu_R$. Then the quark-level amplitude for the Feynman diagrams in Fig.\ref{fig2} can be expressed as
\begin{eqnarray}
\widetilde{\mathcal{M}}_{\phi}^{(b)}
=
i \frac{4 \pi \alpha_s(\mu_R) \lambda_{{\rm color}}}{(k_1 + k_2)^2 - m_{\phi}^2 + i m_{\phi} \Gamma_{\phi}}\,
\epsilon_\mu(k_1) \epsilon_\nu(k_2)\,
\mathcal{C}^{\mu\nu}_{gg\phi}(k_1,\, k_2)\,
\mathcal{C}_{b\bar{b}\phi}
\Big(
\mathcal{F}_a + \mathcal{F}_b + \mathcal{F}_c + \mathcal{F}_d
\Big),
\end{eqnarray}
where $\mathcal{F}_a$, $\mathcal{F}_b$, $\mathcal{F}_c$ and $\mathcal{F}_d$ are given by
\begin{eqnarray}
&&
\mathcal{F}_a
=
\dfrac{1}{(p_2 + q_1)^2}\,
{\rm Tr}
\biggl[
v(p_2) \bar{u}(p_1) \gamma^{\alpha}
\dfrac{1}{(\slashed{k}_1 + \slashed{k}_2) - \slashed{q}_2 - m_b}
v(q_2) \bar{u}(q_1) \gamma_{\alpha}
\biggr],
\\
&&
\mathcal{F}_b
=
\dfrac{1}{(p_2 + q_1)^2}\,
{\rm Tr}
\biggl[
v(p_2) \bar{u}(p_1)
\dfrac{1}{\slashed{p}_1 - (\slashed{k}_1 + \slashed{k}_2) - m_b} \gamma^{\alpha}
v(q_2) \bar{u}(q_1) \gamma_{\alpha}
\biggr],
\\
&&
\mathcal{F}_c
=
\mathcal{F}_b\,\Big|_{p_1 \leftrightarrow q_1,~ p_2 \leftrightarrow q_2}\,,
\\
&&
\mathcal{F}_d
=
\mathcal{F}_a\,\Big|_{p_1 \leftrightarrow q_1,~ p_2 \leftrightarrow q_2}\,.
\end{eqnarray}
Within the framework of NRQCD \cite{Hao:2006nf}, the hadron-level amplitudes $\mathcal{M}_{\phi}^{(g)}$ and $\mathcal{M}_{\phi}^{(b)}$ can be obtained from the quark-level amplitudes $\widetilde{\mathcal{M}}_{\phi}^{(g)}$ and $\widetilde{\mathcal{M}}_{\phi}^{(b)}$, respectively, by performing the following replacement:
\begin{eqnarray}
&&
v(p_2)\bar{u}(p_1)
~~\longrightarrow~~
\dfrac{1}{2\sqrt{2}} \, \rlap/\epsilon^*_{\Upsilon} \big( \rlap/p + m_{\Upsilon} \big)
\dfrac{1}{\sqrt{m_b}}
\Psi_{\Upsilon}(0) \dfrac{1}{\sqrt{N_c}}
\\
&&
v(q_2)\bar{u}(q_1)
~~\longrightarrow~~
\dfrac{1}{2\sqrt{2}} \, \rlap/\epsilon^*_{\Upsilon} \big( \rlap/q + m_{\Upsilon} \big)
\dfrac{1}{\sqrt{m_b}}
\Psi_{\Upsilon}(0) \dfrac{1}{\sqrt{N_c}}
\\
&&
p_1 = p_2 = \dfrac{p}{2}
\\
&&
q_1 = q_2 = \dfrac{q}{2}
\end{eqnarray}

\par
The analytic expression of the SM amplitude for $gg \rightarrow \Upsilon\Upsilon$ (i.e., $\mathcal{M}_{SM}$)  can be obtained analogously, but is not presented here since it is too tedious. Through the standard manipulations, we obtain the cross section $\hat{\sigma}[gg \rightarrow \Upsilon\Upsilon]$ (Eq.(\ref{sigmahat})), and then a convolution with the gluon distribution function results in the cross section for $pp \rightarrow \Upsilon\Upsilon$. In next section we will show our numerical results clearly.

\section{Numerical results}
\par
S. Durgut reported a peak in the invariant mass distribution of $\Upsilon l^+l^-$ at the energy of $18.4~ {\rm GeV}$ \cite{Durgut,phdpaper}. A naive conjecture suggests that the peak at $M_{\Upsilon l^+l^-} \sim 18.4~ {\rm GeV}$ is induced by a BSM Higgs-like boson. Our goal is to check if this scenario works. In this work, the event samples are generated by using {\sc FormCalc} \cite{Hahn:2016ebn} package based on the Monte Carlo technique. In the numerical calculation, the mass of the BSM Higgs-like boson is set as $18.4~ {\rm GeV}$, and thus we denote this Higgs-like boson as $\phi(18.4)$. The factorization and renormalization scales are set to the transverse energy of the final-state $\Upsilon$, i.e., $\mu_F = \mu_R = \sqrt{m_{\Upsilon}^2 + p_{T,\Upsilon}^{\,2}}$. The masses of $b$-quark and $\Upsilon$ are taken as $m_b = 4.73~ {\rm GeV}$ and $m_{\Upsilon} = 9.46~ {\rm GeV}$ \cite{Tanabashi:2018oca}. Within the framework of NRQCD, the zero point wave function of $\Upsilon$ and the branching ratios of $\Upsilon$ to $\mu^+\mu^-$ and $e^+e^-$ are taken as $\Psi_{\Upsilon}^2(0) = 0.391~ {\rm GeV}^3$ \cite{Li:2009ug,Quigg:1977dd,Eichten:1994gt,Eichten:1995ch}, $Br(\Upsilon \rightarrow \mu^+\mu^-) = 2.48\%$ and $Br(\Upsilon \rightarrow e^+e^-) = 2.38\%$ \cite{Tanabashi:2018oca}. The gluon distribution function and the strong coupling constant $\alpha_s$ are adopted from {\sc CT14LO} \cite{Schmidt:2015zda}.

\par
\begin{figure}[htbp]
\begin{center}
\includegraphics[width=0.6\textwidth]{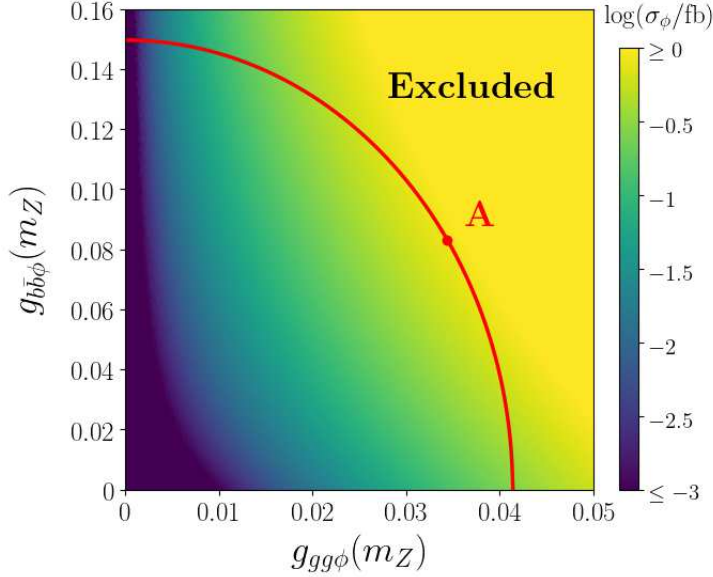}
\caption{Dependence of $\sigma_{\phi}[pp \rightarrow \Upsilon \mu^+\mu^-]$ on the effective coupling constants $g_{gg\phi}$ and $g_{b\bar{b}\phi}$ at the $8~ {\rm TeV}$ LHC. The parameter space region above the red line is excluded by the experimental constraint of $\Gamma[\phi(18.4) \rightarrow gg]+ \Gamma[\phi(18.4) \rightarrow b\bar b] < \Gamma[\phi(18.4) \rightarrow all] \simeq 35~ {\rm MeV}$.}
\label{fig3}
\end{center}
\end{figure}

The dependence of the cross section $\sigma_{\phi}[pp \rightarrow \Upsilon\mu^+\mu^-]$, defined by $\big| \mathcal{M}_{\phi} = \mathcal{M}_{\phi}^{(g)} + \mathcal{M}_{\phi}^{(b)} \big|^2$, on the effective coupling constants $g_{gg\phi}$ and $g_{b\bar{b}\phi}$ at the $8~ {\rm TeV}$ LHC is shown in Fig.\ref{fig3}. Different colors of the points in Fig.\ref{fig3} represent different values of $\sigma_{\phi}$. The parameter space region above the red line is excluded by the experimental constraint from the decay width of $\phi(18.4)$, i.e.,
\begin{eqnarray}
\Gamma[\phi(18.4) \rightarrow gg]+ \Gamma[\phi(18.4) \rightarrow b\bar b] < \Gamma[\phi(18.4) \rightarrow all] \simeq 35~ {\rm MeV}.
\end{eqnarray}
In the experimentally allowed region of the parameter space, the signal cross section $\sigma_{\phi}$ reaches its maximum at the parameter point ${\rm A} = (0.0344,\, 0.0829)$,
\begin{eqnarray}
\sigma_{\phi}[ pp \rightarrow \Upsilon\mu^+\mu^- \, @ ~ 8~ {\rm TeV} ]\Big|_{{\rm A}}
=
0.853~ {\rm fb}.
\end{eqnarray}
It is obvious that $\sigma_{\phi}[ pp \rightarrow \Upsilon\mu^+\mu^- \, @ ~ 8~ {\rm TeV} ] < 0.853~ {\rm fb}$ in the whole experimentally allowed parameter space region. The purpose of this study is to investigate whether the existence of a BSM Higgs-like boson can fit the enhancement observed by the CMS collaboration. Therefore, we give preference to the parameter point A in the following discussion.
%%%%%%%%%%%%%%%%%%%%% Fig. 3 %%%%%%%%%%%%%%%%%%%%%%%%
%%%%%%%%%%%%%%%%%%%%%%%%%%%%%%%%%%%%%%%%%%%%%%%%%%%%

\par
\begin{figure}[htbp]
\begin{center}
\includegraphics[width=0.6\textwidth]{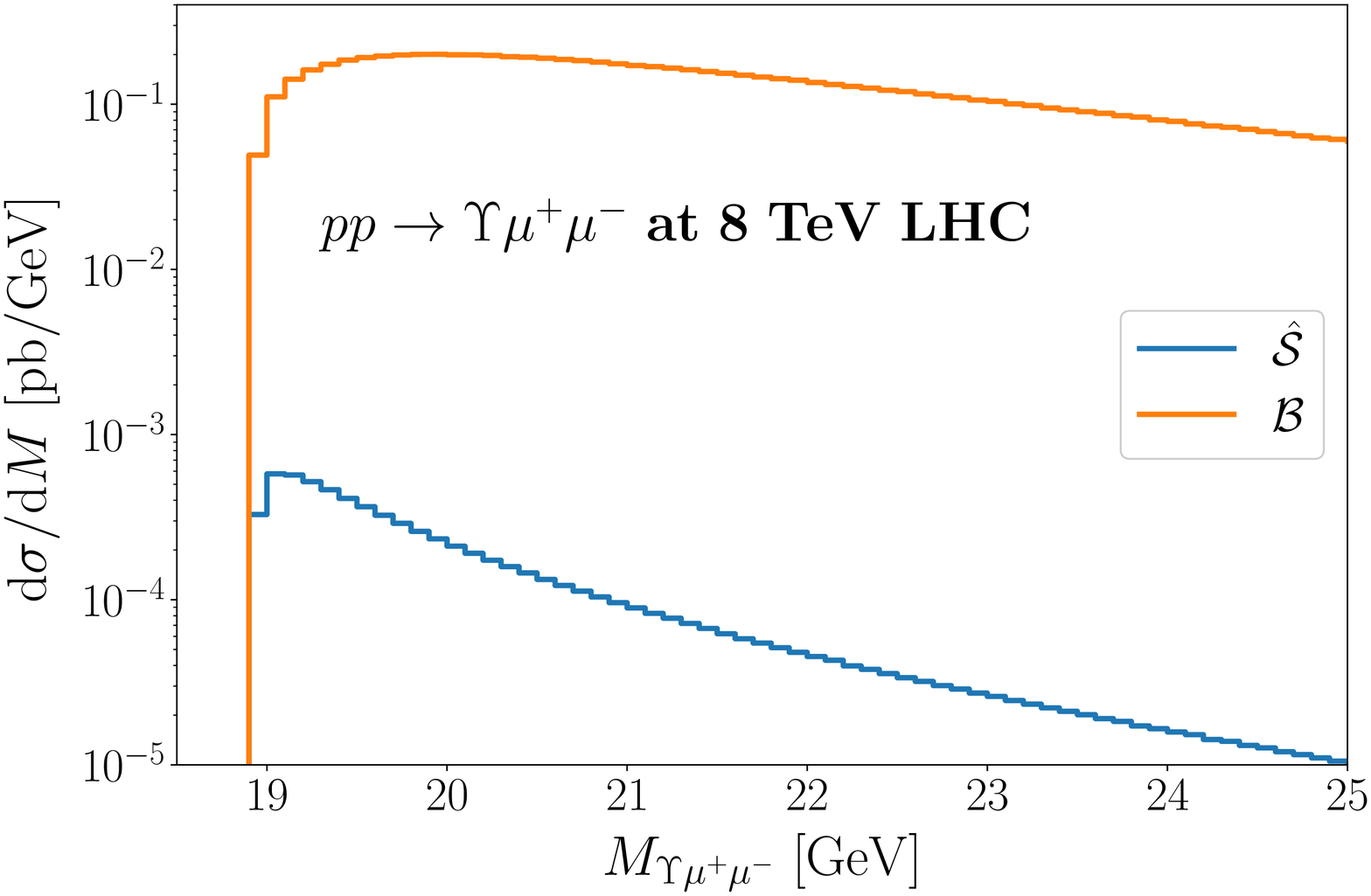}
\includegraphics[width=0.6\textwidth]{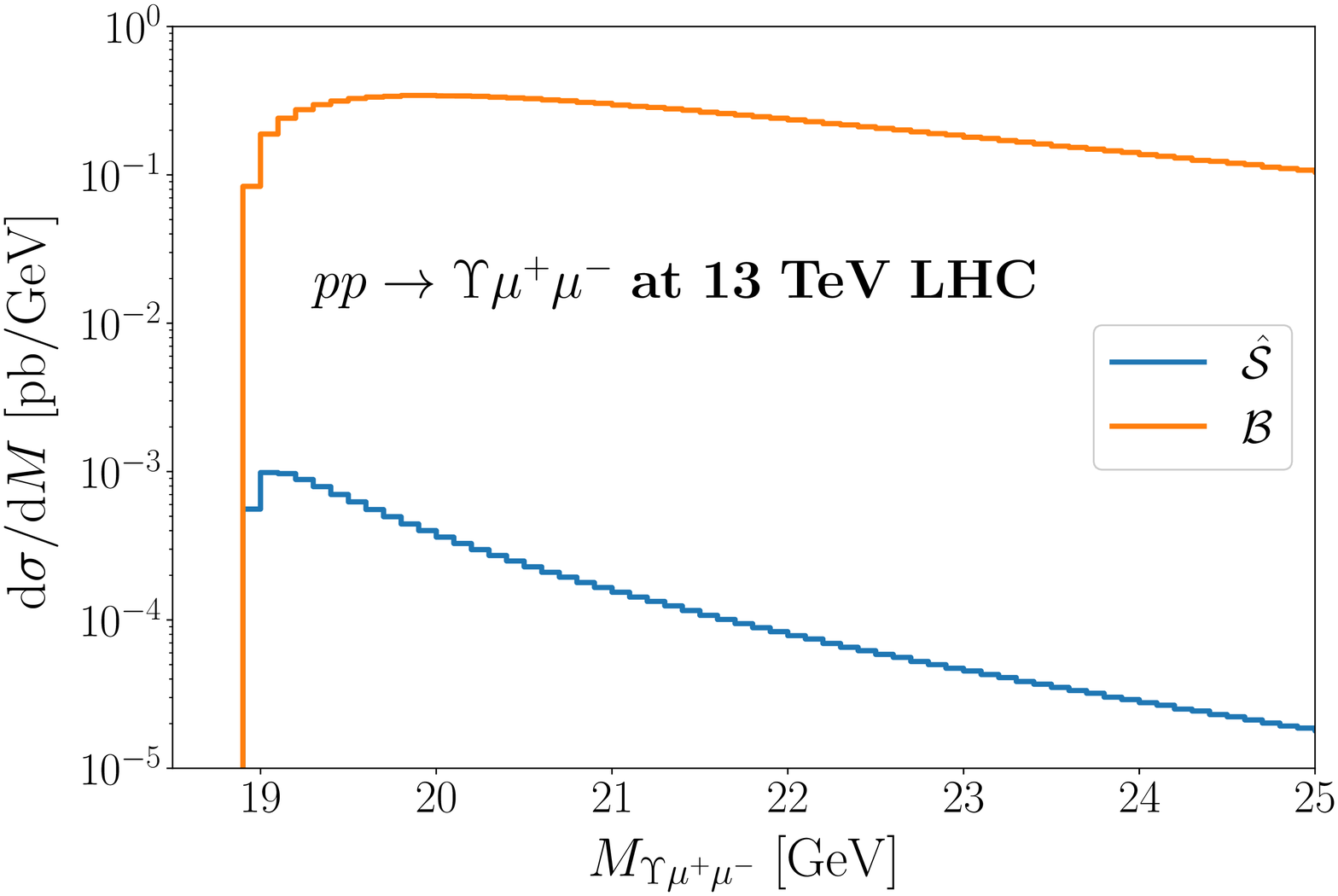}
\caption{$\Upsilon\mu^+\mu^-$ invariant mass distributions for $pp \rightarrow \Upsilon\mu^+\mu^-$  at the $8$ and $13~ {\rm TeV}$ LHC. $B$ stands for the SM background and $\hat{S}$ represents the new physics signal induced by $\phi(18.4)$ without interference effect, respectively. The decay width and the effective coupling constants of $\phi(18.4)$ are taken as $\Gamma_{\phi} = 35~ {\rm MeV}$, $g_{gg\phi}(m_Z) = 0.0344$ and $g_{b\bar{b}\phi}(m_Z) = 0.0829$.}
\label{fig4}
\end{center}
\end{figure}

The integrated cross sections and invariant mass spectra of $\Upsilon \mu^+\mu^-$ for $pp \rightarrow \Upsilon \mu^+\mu^-$ at the $8$ and $13~ {\rm TeV}$ LHC are provided in Tab.\ref{tab1} and Fig.\ref{fig4}, respectively. The contributions from $\left| \mathcal{M}_{SM} \right|^2$, $\left| \mathcal{M}_{\phi} \right|^2 + 2 {\rm Re}\big( \mathcal{M}_{SM}^{\dag} \mathcal{M}_{\phi} \big)$ and $\left| \mathcal{M}_{\phi} \right|^2$, which are regarded as the SM background and the new physics signals induced by the BSM Higgs-like boson with and without interference effect, are provided separately, and labeled with $B$, $S$ and $\hat{S}$ respectively. Table \ref{tab1} clearly shows that the interference between the BSM amplitude induced by $\phi(18.4)$ and the SM amplitude for $pp \rightarrow \Upsilon \mu^+\mu^-$ is negative, and thus reduces the new physics signal induced by $\phi(18.4)$ in $pp \rightarrow \Upsilon \mu^+\mu^-$ production. One can notice that the contribution of $\phi(18.4)$ at colliding energy between two gluons being below $2 m_{\Upsilon}$ is almost zero, but would jump up at $\sqrt{s} = 2 m_{\Upsilon}$. It is a standard threshold effect. One characteristic of the phenomenon is the observed ``peak" is not in the symmetric Gaussian form. Anyhow, even though we suppose existence of a BSM Higgs-like boson which may decay into $\Upsilon\mu^+\mu^-$, it is impossible to induce a peak at $18.4~ {\rm GeV}$ at all. What's more, the extremely narrow peak at $M_{\Upsilon \mu^+\mu^-} \sim 18.4~ {\rm GeV}$ in the invariant mass spectrum of $\Upsilon \mu^+\mu^-$ in the D-report ($\Gamma[\phi(18.4) \rightarrow all] < 35~ {\rm MeV}$) \cite{Durgut,phdpaper} gives a stringent constraint on the effective couplings of the Higgs-like boson $\phi(18.4)$, especially the coupling to two gluons.
%%%%%%%%%%%%%% Table 1 %%%%%%%%%%%%%%%%%%%%%%%%%%%%%
\begin{table}[h]
\begin{center}
\renewcommand\arraystretch{1.8}
\begin{tabular}{cccc}
\hline
\hline
~~$\sqrt{s}$~ [TeV]~~ & ~~$\sigma_{\hat{S}}$~ [fb]~~ & ~~$\sigma_{S}$~ [fb]~~ & ~~$\sigma_{B}$~ [pb]~~  \\
\hline
$8$   &  $0.853$ &  $-0.637$ &  $1.041$   \\
$13$  &  $1.464$ &  $-1.073$ &  $1.811$   \\
\hline
\hline
\end{tabular}
\caption{Integrated cross sections for $pp \rightarrow \Upsilon\mu^+\mu^-$ at the $8$ and $13~ {\rm TeV}$ LHC. $B$ stands for the SM background, $S$ and $\hat{S}$ represent the new physics signals induced by $\phi(18.4)$ with and without interference effect, respectively. The decay width and the effective coupling constants of $\phi(18.4)$ are taken as $\Gamma_{\phi} = 35~ {\rm MeV}$, $g_{gg\phi}(m_Z) = 0.0344$ and $g_{b\bar{b}\phi}(m_Z) = 0.0829$.}
\label{tab1}
\end{center}
\end{table}
%%%%%%%%%%%%%%%%%%%%%%%%%%%%%%%%%%%%%%%%%%%%%%%%%%%%%
%%%%%%%%%%%%%%%%%%%%% Fig. 4 %%%%%%%%%%%%%%%%%%%%%%%%
%%%%%%%%%%%%%%%%%%%%%%%%%%%%%%%%%%%%%%%%%%%%%%%%%%%%

\par
In Tab.\ref{tab2}, we present the production cross sections for both signal and SM background for $\Upsilon e^+e^-$ final state. We can see clearly that the numerical results for the $pp \rightarrow \Upsilon e^+e^-$ channel are almost the same as those for $pp \rightarrow \Upsilon \mu^+\mu^-$ due to the lepton universality ($Br(\Upsilon \rightarrow e^+e^-) \simeq Br(\Upsilon \rightarrow \mu^+\mu^-)$). Thus, we do not provide the invariant mass distribution of $\Upsilon e^+e^-$ in this section.
%%%%%%%%%%%%%% Table 2 %%%%%%%%%%%%%%%%%%%%%%%%%%%%%
\begin{table}[h]
\begin{center}
\renewcommand\arraystretch{1.8}
\begin{tabular}{cccc}
\hline
\hline
~~$\sqrt{s}$~ [TeV]~~ & ~~$\sigma_{\hat{S}}$~ [fb]~~ & ~~$\sigma_{S}$~ [fb]~~ & ~~$\sigma_{B}$~ [pb]~~  \\
\hline
$8$   &  $0.819$ &  $-0.611$ &  $0.999$   \\
$13$  &  $1.405$ &  $-1.030$ &  $1.738$   \\
\hline
\hline
\end{tabular}
\caption{The same as Tab.\ref{tab1} but for $\Upsilon e^+e^-$ final state.}
\label{tab2}
\end{center}
\end{table}
%%%%%%%%%%%%%%%%%%%%%%%%%%%%%%%%%%%%%%%%%%%%%%%%%%%%%

\section{Discussions and conclusion}
\par
Based on the data of the Run I of LHC at $7$ and $8~ {\rm TeV}$, we investigate  the origin of the peak at $18.4~ {\rm GeV}$ in the invariant mass spectrum of $\Upsilon l^+l^-$ newly reported in Refs.\cite{Durgut,phdpaper}. We postulate it to be a $0^{++}$ BSM Higgs-like boson, and by the anzatz calculate the production rate of $\Upsilon l^+l^-$ via gluon-gluon fusion at the LHC and discuss the effect of this BSM Higgs-like boson. In our calculations, we assume the effective couplings of the $0^{++}$ BSM Higgs-like boson to the gluons and bottom quarks have the same evolution behaviour as the corresponding ones of the SM Higgs boson.

\par
For the peak observed in the invariant mass spectrum of $\Upsilon l^+l^-$ at $18.4~ {\rm GeV}$ whose width was not accurately fixed yet, the situation might imply that the peak corresponds to a BSM Higgs-like boson which decays into $\Upsilon\Upsilon^*$ and later turns into $\Upsilon l^+l^-$ and eventually goes to the four-lepton final state. The peak position is located at $18.4~ {\rm GeV}$ which is lower than the threshold value of $2 m_{\Upsilon}$, so that it impossibly directly decays into a real $\Upsilon$ pair if we do not consider its width. If it possesses a relatively large width whose edge covers the region of $2 m_{\Upsilon}$, it may result in an asymmetric peak at $M_{\Upsilon l^+l^-} \simeq 2 m_{\Upsilon}$ in the invariant mass spectrum of $\Upsilon l^+l^-$ via the threshold effect. We carefully analyze the possibility and our numerical results (Fig.\ref{fig4}) assure  that there cannot exist an even-not-very apparent asymmetric peak above $2 m_{\Upsilon}$. Moreover, the contribution of the new Higgs-like boson to the portal $\Upsilon l^+l^-$ would interfere with the SM contribution and accurate measurements may detect the variation. But all the numerical results do not manifest an appearance of a peak at $18.4~ {\rm GeV}$.

\par
From Tab.\ref{tab1}, Tab.\ref{tab2} and Fig.\ref{fig4} we can notice that if the coupling constants are small and the width of the supposed Higgs-like boson is narrow ($\Gamma[\phi(18.4) \rightarrow all] < 35~ {\rm MeV}$), the cross section is $\mathcal{O}(1~{\rm fb})$, such a small cross section cannot be experimentally observed by the present facilities. Since the effect of the BSM Higgs-like boson $\phi(18.4)$ cannot be detected in $\Upsilon l^+l^-$ final state at the parameter point ${\rm A} = (0.0344,\, 0.0829)$, the whole parameter space region allowed by the constraint of $\Gamma[\phi(18.4) \rightarrow gg]+ \Gamma[\phi(18.4) \rightarrow b\bar b] < \Gamma[\phi(18.4) \rightarrow all] \simeq 35~ {\rm MeV}$ is entirely excluded by the peak observed in the $\Upsilon l^+l^-$ invariant mass spectrum at $18.4~ {\rm GeV}$. If this scenario is true, we would conclude that the peak at $18.4~{\rm GeV}$ does not correspond to a $0^{++}$ BSM Higgs-like boson, but something else.

\par
All of our estimates are based on the experimental results reported in Refs.\cite{Durgut,phdpaper}. Our numerical results decide that the peak may not corresponds to a $0^{++}$ BSM Higgs-like boson ($18.4~ {\rm GeV}$). Definitely much more accurate measurements which will be carried out at future high energy facilities (including the updated LHC) will give more information about this peak.

\par
By our assumption, the observed peak is a BSM Higgs-like boson, if it is true, it would set a scale for the BSM and the significance is obvious. Indeed, for the peak appearing at the invariant mass spectrum of $\Upsilon l^+l^-$, Refs.\cite{Karliner:2016zzc,Esposito:2018cwh,Becchi:2020mjz} consider it to be a composite of $bb\bar{b}\bar{b}$, but all their study show that the decay width are too small to be currently observed at the LHC. The observation is important and following the data, the theoretical interpretation can be made. Since it implies new understanding on new physics beyond the SM and sets a new scale, obviously, the study along this line cannot be neglected. We hope the experimentalists of high energy physics to continue the investigation on the peak by more accurate measurement and analysis. The conclusion would greatly help theorists making a definite judgement to verify the validity of our ansatz or negate it.

\par
Now let us make a brief summary and draw our conclusion (so far, but by no means for the future). In this work we are trying to investigate whether the enhancement observed at LHC is a structureless BSM boson. If it indeed is, it can contribute to the process of $pp \rightarrow \Upsilon l^+l^-$, but how it behaves, can it result in a peak at the invariant mass spectrum of $\Upsilon l^+l^-$, in other words, does it induce the peak at $18.4~ {\rm GeV}$ reported in Refs.\cite{Durgut,phdpaper}? It demands a clear answer. Even though a BSM boson $\phi$ exists and possesses a certain width, an inequality $m_{\phi}+ \Gamma_\phi < 2 m_{\Upsilon}$ holds. Our explicit computation  indicates that $\phi$ as an on-shell real particle may not directly contribute to $pp \rightarrow \phi \rightarrow \Upsilon\Upsilon^* \rightarrow \Upsilon l^+l^-$. Thus even though a BSM Higgs-like boson $\phi$ of $18.4~ {\rm GeV}$ exists and may contribute to $pp \rightarrow \Upsilon l^+l^-$, the sizable rate only occurs above the threshold of 2$m_{\Upsilon}$. But then $\phi$ must be off-shell (or contributes via $t$- and $u$-channels), therefore our conclusion is that the experimentally observed peak located at $18.4~ {\rm GeV}$ with a narrow width does not correspond to a BSM structureless Higgs-like boson. The  peak of $18.4~ {\rm GeV}$ must originate from other mechanism and its appearance cannot be a signature of existence of BSM as expected.

\vskip 10mm
\par
\noindent{\large\bf ACKNOWLEDGMENTS} \\
This work is supported in part by the National Natural Science Foundation of China (Grants No. 11675082, 11735010, 11775211, 11535002, 11805160, 11747040, 11375128, the Natural Science Foundation of Jiangsu Province of China (Grant No. BK20170247), Special Grant of the Xuzhou University of Technology (No. XKY2016211, XKY2017215, XKY2018221),  and the CAS Center for Excellence in Particle Physics (CCEPP).

\end{document}